\documentclass[aps,pre,reprint]{revtex4-1}

\usepackage{graphicx}
\usepackage{epstopdf}
\usepackage{dcolumn}
\usepackage{bm}
\usepackage{amsmath}
\usepackage{amssymb}
\usepackage{wasysym}
\usepackage{color}
\usepackage{amssymb}
\usepackage{dsfont}
\usepackage{subfigure}

\newcommand{\dif}{\mathrm{d}}

\begin{document}

\title{Existence of negative differential thermal conductance in one-dimensional diffusive thermal transport}

\author{Jiuning Hu$^{1,2}$}
\email[]{hu49@purdue.edu}
\author{Yong P. Chen$^{3,2,1}$}

\affiliation{$^1$School of Electrical and Computer Engineering, Purdue University, West Lafayette, Indiana 47907,USA\\
$^2$Birck Nanotechnology Center, Purdue University, West Lafayette, Indiana 47907,USA\\
$^3$Department of Physics, Purdue University, West Lafayette, Indiana 47907,USA}

\begin{abstract}
We show that in a finite one-dimensional (1D) system with diffusive thermal transport described by the Fourier's law, negative differential thermal conductance (NDTC) cannot occur when the temperature at one end is fixed. We demonstrate that NDTC in this case requires the presence of junction(s) with temperature dependent thermal contact resistance (TCR). We derive a necessary and sufficient condition for the existence of NDTC in terms of the properties of the TCR for systems with a single junction. We show that under certain circumstances we even could have infinite (negative or positive) differential thermal conductance in the presence of the TCR. Our predictions provide theoretical basis for constructing NDTC-based devices, such as thermal amplifiers, oscillators and logic devices.
\end{abstract}


\maketitle

In recent years, nonlinear thermal transport, particularly in low dimensional systems, is of significant interest from both fundamental and practical perspectives \cite{Lepri03,RevModPhys.84.1045}. For example, thermal rectification has been experimentally and theoretically studied in many nanostructures \cite{PhysRevLett.88.094302,PhysRevLett.97.094301,Li04,PhysRevLett.98.104302,Peyrard06,Chang06,Yang07,Hu09,Pereira10}  and heterogeneous bulk materials \cite{Sawaki11,Go10,Dames09}. Negative differential thermal conductance (NDTC), an unusual thermal transport phenomenon where the heat current across a thermal conductor 
decreases when the temperature bias increases, is an essential element for the construction of thermal transistors \cite{Li06} and thermal logic \cite{Wang07},  and is shown to exist in many non-linear one-dimensional (1D) systems \cite{Li06,Yang07,He09,Zhong09,Shao09,He10,Pereira10,Ai11,Ai11b,Hu11,Shao11} and vacuum gaps \cite{zhu044104}. Many mechanisms such as nonlinear interactions \cite{Segal06}, molecular anharmonicity \cite{PRE.80.041103,Ai11,Ai11b,Pereira10}, interplay between the thermal driving force and the thermal (boundary) conductance \cite{He09,He10,Zhong09,Hu11}, thermal interfaces \cite{Li06,Shao09,He09} and others \cite{zhu044104} have been proposed to explain the existence of NDTC. Interestingly, several numerical studies  \cite{Shao09,Zhong09,He10,Hu11} have suggested that NDTC may vanish as the system length becomes large (approaching diffusive thermal transport). However, it has not been definitely answered whether NDTC universally vanishes for diffusive thermal transport. Besides, the role played by thermal interfaces in NDTC has not been well studied. Here, we provide a generic and analytic study of these issues in 1D diffusive thermal transport described by the Fourier's law. We prove that NDTC cannot exist when the temperature at one end is fixed. However, we show that NDTC in this case is still possible if a junction with temperature dependent thermal contact resistance (TCR) is introduced. Unlike previous theories and simulations \cite{Li06,Yang07,He09,Zhong09,Shao09,He10,Pereira10,Ai11,Ai11b,Hu11,Shao11} that dealt with specific toy models which are often difficult to access experimentally, our predictions provide a generic way towards building NDTC-based devices.

We consider a general 1D system in the diffusive thermal transport regime whose thermal conductivity $\kappa(x,T)$ is a function of the coordinate $x$ and the local temperature $T(x)$. The position dependence of the thermal conductivity $\kappa(x,T)$ is explicitly expressed, since the system we consider can have a spatial dependence of structure or composition (e.g., strain or mass gradient). This phenomenological description is valid as long as the mean free path (MFP) of heat carriers is much smaller than the size of the system, where the microscopic details are unimportant. This approach generates analytic results regarding the existence of NDTC, and it is instructive in system design to pursue the applications of NDTC.

\begin{figure}
  \includegraphics[width=0.5\textwidth]{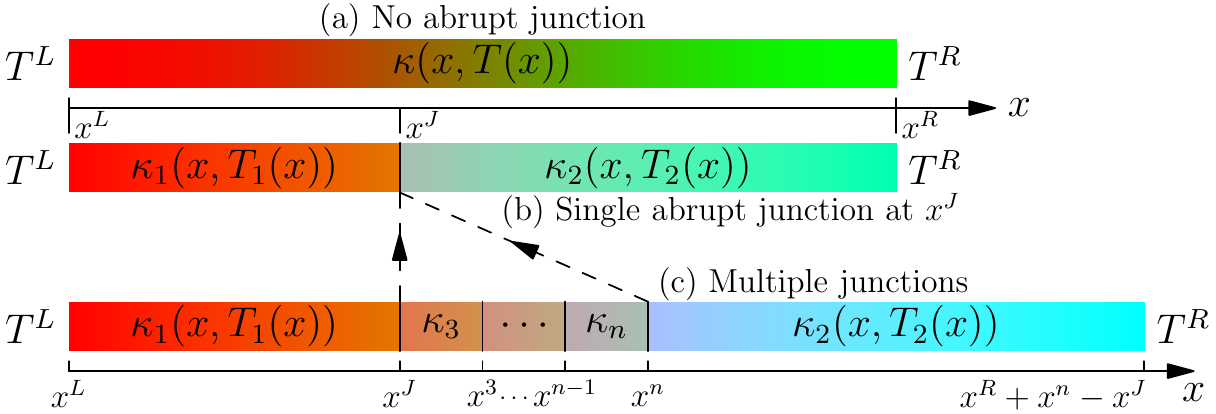}\\
  \caption{Schematics of 1D systems without any junctions (a) and with a single junction (b) and multiple junctions (c). The junctions are indicated by vertical black lines. }
  \label{fig:scheme}
\end{figure}

For a finite 1D system which lies in the coordinate range $[x^L,x^R]$ (Fig.~\ref{fig:scheme}), the local heat current $q(x)$ can be calculated from the Fourier's law:
\begin{equation}
q(x)=-\kappa(x,T)\frac{\dif T}{\dif x}.
\label{heat current}
\end{equation}
For thermal transport without heat sources or sinks, the heat current is conserved and the steady state thermal transport equation reads
\begin{equation}
\frac{\dif}{\dif x}\left(\kappa(x,T)\frac{\dif T}{\dif x}\right)=0.
\label{heat eq}
\end{equation}
Once the temperature at two ends of the system are given, i.e., 
\begin{equation}
T\left(x^{L(R)}\right)=T^{L(R)},
\label{heat boundary}
\end{equation}
the temperature profile $T(x)$ is uniquely \cite{Walter98} determined by Eq.~(\ref{heat eq}) and the boundary conditions (\ref{heat boundary}), and the resulting heat current $q$ (independent of $x$) flowing in the system can be computed from Eq.~(\ref{heat current}).

By applying an infinitesimal variation $\delta T^{L(R)}$ of the boundary temperature at one end, i.e., $T^{L(R)}$ is varied to $T^{L(R)}+\delta T^{L(R)}$ while the temperature $T^{R(L)}$ at the other end is fixed, the resulting temperature profile is varied to $T(x)+\delta T(x)$. 
This temperature profile variation $\delta T(x)$ can induce a variation $\delta q$ of the heat current. We define the differential thermal conductance (DTC) as
\begin{equation}
\begin{aligned}
G\equiv\frac{\delta q}{\delta(T^L-T^R)},
\label{DTC}
\end{aligned}
\end{equation}
or specifically
\begin{equation}
\begin{aligned}
&G=G^L=\frac{\delta q}{\delta T^L},\;\;\;\;\;\text{when}\;T^R\;\text{is fixed; or}\\
&G=G^R=-\frac{\delta q}{\delta T^R},\;\;\text{when}\;T^L\;\text{is fixed}.
\end{aligned}
\label{DTC def}
\end{equation}
In the following we consider the cases with and without the junctions to discuss the existence of NDTC.

\vspace{0.5cm}

\noindent{\textit{Systems without abrupt junctions}, as shown in Fig.~\ref{fig:scheme}(a). In this case, if $T^{R(L)}$ is fixed (without loss of generality, we can assume $T^L>T^R$), we will show that there is no NDTC and the heat current $q$ will increase as the temperature bias increases by increasing (lowering) $T^{L(R)}$ (NDTC could still exist when the temperatures at two ends vary simultaneously \cite{note1}, however, we limit our study in the cases that the temperature at one end is fixed). 

Qualitatively, we first point out that the non-existence of NDTC here is a direct consequence of the uniqueness of the solution of Eq.~(\ref{heat eq}), as we graphically demonstrate in Fig.~\ref{fig:Tprof}(a). Take the case that $T^R$ is fixed as an example. If the temperature $T^L$ is increased to $T^{\prime L}$ ($>T^L$), the temperature profile $T(x)$ (black line) with $T(x^{L(R)})=T^{L(R)}$ and heat current $q$ are changed to $T^\prime(x)$ (red line) with $T^\prime(x^L)=T^{\prime L}$ and $T^\prime(x^R)=T^R$ and heat current $q^\prime$. First of all, we must have $q^\prime\neq q$. Otherwise, the first order differential equation $\frac{\dif T}{\dif x}=-\frac{q}{\kappa(x,T)}$ about $T$ with initial condition $T(x^R)=T^R$ would have non-unique solutions ($T(x)$ and $T^\prime(x)$), which is not allowed. Secondly, $q^\prime$ cannot be smaller than $q$ (proportional to the slope of $T(x)$ at $x^R$), because otherwise there will be an intersection of $T^\prime(x)$ (represented by the dashed line) and $T(x)$ at some $x^I<x^R$. We then must have $T^\prime(x)=T(x)$ in the coordinate range $[x^I,x^R]$ due to the uniqueness of the solution to Eq.~(\ref{heat eq}), and thus $q^\prime=q$ (contradiction). Therefore, we must have $q^\prime>q$, i.e., the heat current monotonically increases with temperature $T^{L}$ when $T^{R}$ is fixed and there is no NDTC. Similar arguments apply to the case when $T^L$ is fixed.

We have derived the analytical expressions for the DTCs as (Appendix A)
\begin{equation}
G^L=J^{-1},\;\;G^R=F(x^R)J^{-1},
\label{GLR}
\end{equation}
where 
\begin{equation}
\begin{aligned}
&F(x)\equiv\exp\left(\int_{x^L}^x\frac{1}{\kappa(x^\prime,T(x^\prime))}\frac{\partial\kappa}{\partial T}\frac{\dif T}{\dif x^\prime}\dif x^\prime\right),\\
&J\equiv\int_{x^L}^{x^R}\frac{F(x^\prime)}{\kappa(x^\prime,T(x^\prime))}\dif x^\prime.
\end{aligned}
\label{FJ}
\end{equation}
Such expressions are useful to calculate the magnitude of DTCs from the temperature profile without the needs to know directly the heat current and its variation (as in Eq.~(\ref{DTC})). They also directly prove the non-existence of NDTC here: since $F(x)$ and $\kappa$ are positive, we have $G^L>0$ and $G^R>0$.

\begin{figure}%
\centering
\subfigure{\includegraphics[width=0.5\textwidth]{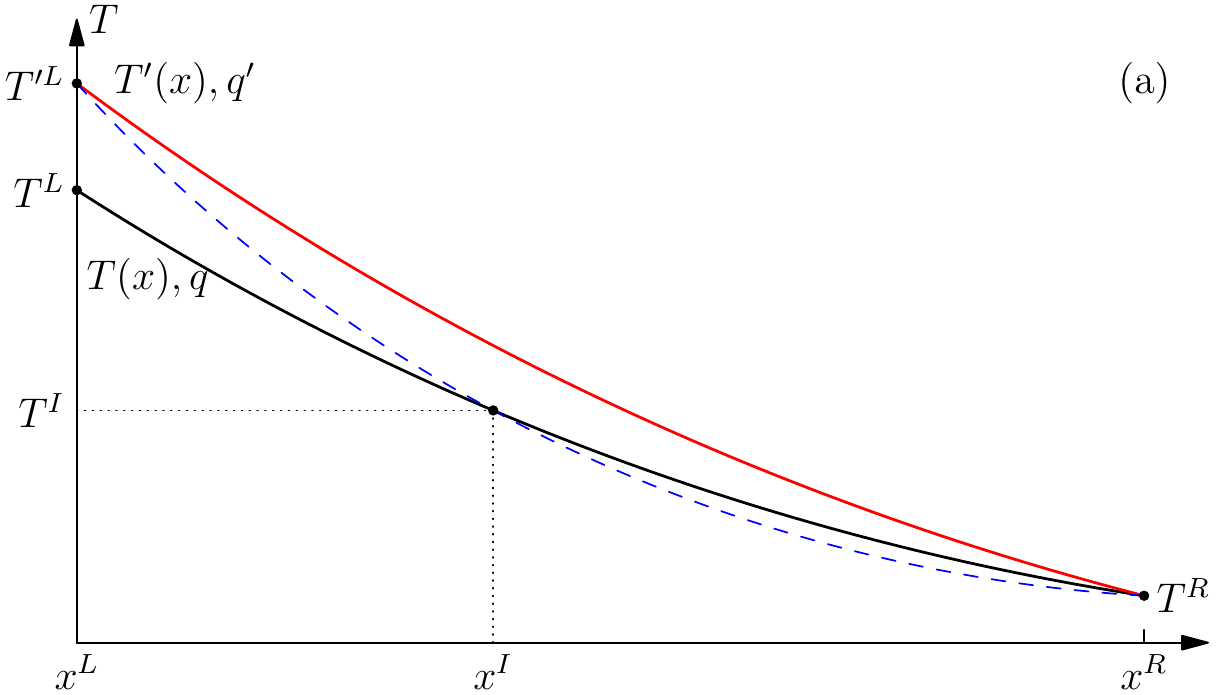}}\\
\subfigure{\includegraphics[width=0.5\textwidth]{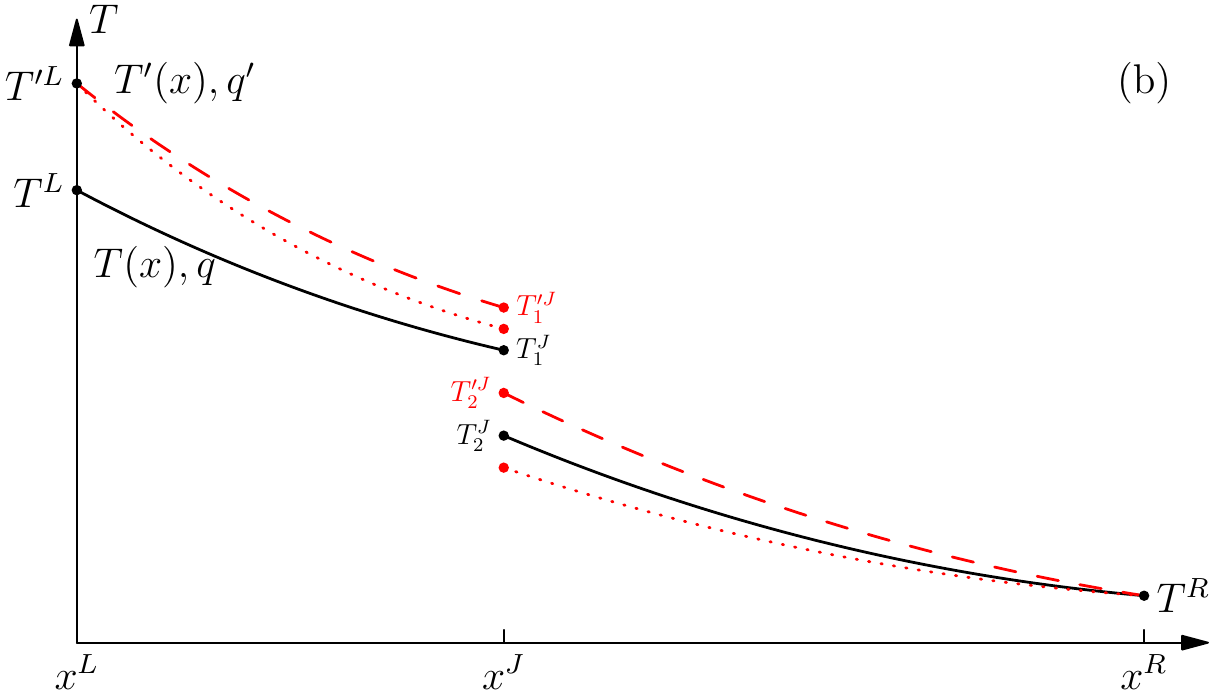}}
\caption{A schematic example of temperature profiles of systems (a) without any junctions and (b) with a single junction at $x^J$ when the temperature at $x^R$ is fixed and the temperature at $x^L$ is increased from $T^L$ to $T^{\prime L}$. The dotted lines in (b) would give rise to NDTC.}
\label{fig:Tprof}
\end{figure}


\vspace{0.5cm}

\noindent\textit{Systems with a single abrupt junction}, as shown in Fig.~\ref{fig:scheme}(b). We assume that the abrupt junction is located at $x^J\in(x^L,x^R)$. The system can be considered as two subsystems without any abrupt junctions coupled together at $x^J$. We suppose that the subsystems in $[x^L,x^J]$ and $[x^J,x^R]$ have thermal conductivity $\kappa_1(x,T_1)$ with temperature profile $T_1(x)$ and $\kappa_2(x,T_2)$ with temperature profile $T_2(x)$ respectively. We denote
\begin{equation}
\begin{aligned}
&T_1^J\equiv T_1(x^J),\;T_2^J\equiv T_2(x^J),\\&T^L\equiv T_1(x^L),\;T^R\equiv T_2(x^R).
\end{aligned}
\label{boundary condition coupled}
\end{equation}
At the junction, the two subsystems are coupled through a thermal contact resistance (TCR) $R^J$, defined such that the heat current $q$ flowing through the system satisfies \cite{note2}
\begin{equation}
T_1^J-T_2^J=qR^J(T_1^J,T_2^J).
\label{junction boundary condition}
\end{equation}
We have generally assumed that the TCR depend on two temperatures $T_1^J$ and $T_2^J$. The TCR $R^J(T_1^J,T_2^J)$ provides the complete characterizations of the junction at the phenomenological level.

The possibility of NDTC here can also be interpreted graphically. Again, take the case that $T^R$ is fixed as an example. If the temperature $T^L$ ($>T^R$) is increased to $T^{\prime L}$ ($>T^L$), the temperature profile $T(x)$ with $T(x^{L(R)})=T^{L(R)}$, junction temperatures $T_{1,2}^J$ and heat current $q$ are changed to $T^\prime(x)$ with $T^\prime(x^L)=T^{\prime L}$ and $T^\prime(x^R)=T^R$, junction temperatures $T_{1,2}^{\prime J}$ and heat current $q^\prime$, as illustrated in Fig.~\ref{fig:Tprof}(b). Because of the discontinuous jump of temperature at $x^J$, we could have $q^\prime>q$ if $T_2^{\prime J}>T_2^{J}$ (red dashed line) or $q^\prime<q$ if $T_2^{\prime J}<T_2^{J}$ (red dotted line) where the latter situation gives rise to NDTC. We derive the conditions of the existence of NDTC in more detail below.

Analytically, the DTCs are (Appendix B)
\begin{equation}
\begin{aligned}
G^L\!\!=\!\!\left(\! 1\!-\!q\frac{\partial R^J}{\partial T_1^J}\!\right)\!\!\frac{F_1(x^L)}{R^D},
G^R\!\!=\!\!\left(\! 1\!+\!q\frac{\partial R^J}{\partial T_2^J}\!\right)\!\!\frac{F_2(x^R)}{R^D},
\end{aligned}
\label{GLR coupled}
\end{equation}
with
\begin{equation}
F_i(x)\!\equiv\!\exp\!\!\left(\int_{x^J}^x\!\frac{1}{\kappa_i(x^\prime\!,T_i(x^\prime))}\frac{\partial\kappa_i}{\partial T_i}\frac{\dif T_i}{\dif x^\prime}\dif x^\prime\!\right)\!,i\!=\!1,2,
\label{Fi}
\end{equation}
defined on $[x^L,x^J]$ and $[x^J,x^R]$ respectively, and
\begin{equation}
R^D\equiv R^J+\displaystyle\left(1-q\frac{\partial R^J}{\partial T_1^J}\right)J_1+\left(1+q\frac{\partial R^J}{\partial T_2^J}\right)J_2
\label{RD}
\end{equation}
where
\begin{equation}
J_1\!\!\equiv\!\!\!\int_{x^L}^{x^J}\!\!\!\frac{F_1(x^\prime)}{\kappa_1\!(x^\prime\!,T_1\!(x^\prime))}\dif x^\prime,
J_2\!\!\equiv\!\!\!\int_{x^J}^{x^R}\!\!\!\frac{F_2(x^\prime)}{\kappa_2\!(x^\prime\!,T_2\!(x^\prime))}\dif x^\prime.
\label{J1J2}
\end{equation}

If the TCR is independent of the junction temperatures $T_1^J$ and $T_2^J$, the partial derivatives of $R^J$ in Eqs.~(\ref{GLR coupled}) and (\ref{RD}) vanishes, and since $F_i$ in Eq.~(\ref{Fi}) and $J_i$ in Eq.~(\ref{J1J2}) are positive, thus $G^{L(R)}>0$ and there is no NDTC. Therefore, a temperature dependent TCR is necessary for the existence of NDTC. However, as we will see, it is not a sufficient condition.

In the presence of the temperature dependence of TCR, we pick a $T_B$ such that $T_B^{-1}>\vert R^{-1}\partial R^J/\partial T_{1,2}^J\vert$, where $R=(T^L-T^R)/q$ is the thermal resistance of the whole system including the TCR. Inside the regime defined by $\vert T^L-T^R\vert<T_B$ in the $(T^L,T^R)$ quarter plane we have $\vert q\partial R^J/\partial T^J_{1,2}\vert<1$ and subsequently $R^D>0$ and $G^{L(R)}>0$: no NDTC is displayed in this low bias regime. Therefore, a temperature bias exceeding $T_B$ is required to observe NDTC, confirming that NDTC is a \emph{nonlinear} thermal transport phenomenon.


As the temperature bias ($\vert T^L-T^R\vert$) increases beyond $T_B$, $G^{L(R)}$ could possibly be negative, leading to NDTC. We denote the dimensionless quantities
\begin{equation}
\begin{aligned}
&X\equiv\left(1-q\frac{\partial R^J}{\partial T_1^J}\right)\frac{J_1}{R^J}=\frac{\partial q}{\partial T_1^J}J_1,\\
&Y\equiv\left(1+q\frac{\partial R^J}{\partial T_2^J}\right)\frac{J_2}{R^J}=-\frac{\partial q}{\partial T_2^J}J_2
\end{aligned}
\label{xy}
\end{equation}
such that
\begin{equation}
G^L=\frac{X}{1\!+\!X\!+\!Y}\frac{F_1(x^L)}{J_1},G^R=\frac{Y}{1\!+\!X\!+\!Y}\frac{F_2(x^R)}{J_2}.
\label{Gxy}
\end{equation}
Now there exists NDTC if and only if at least one of $X$ and $Y$ is negative, which means that at least one of $\partial q/\partial T_1^J$ and $-\partial q/\partial T_2^J$ is negative \cite{note3}. We refer to such junctions as those with \emph{intrinsic junction NDTC} which is now necessary and sufficient for NDTC to occur. Thus, the existence of NDTC in systems with a single abrupt junction is uniquely determined by the properties of the TCRs, regardless of the properties of the system away from the junction.

Furthermore, we can formulate the existence of NDTC on the $X$-$Y$ plane: find out the points on the plane that correspond to negative $G^L$ or $G^R$. NDTC exists inside the shaded areas (A, B and C in Fig.~\ref{fig:xy}), not including the boundaries labelled by the thick solid black and red dashed lines. Note that we have $G^L<0$ and $G^R=0$ on the thin dotted line ($-1<X<0$, $Y=0$) while $G^L=0$ and $G^R<0$ on the thin dash-dotted line ($X=0$, $-1<Y<0$). We have $R^D=0$ and thus infinite ($\pm\infty$) DTCs on the thick red dashed line \cite{note4}. For the points in the shaded areas and close to the thick red dashed line, we can have very large magnitude of NDTC, useful to design sensitive detectors for temperature fluctuations.

\begin{figure}
  \includegraphics[width=0.45\textwidth]{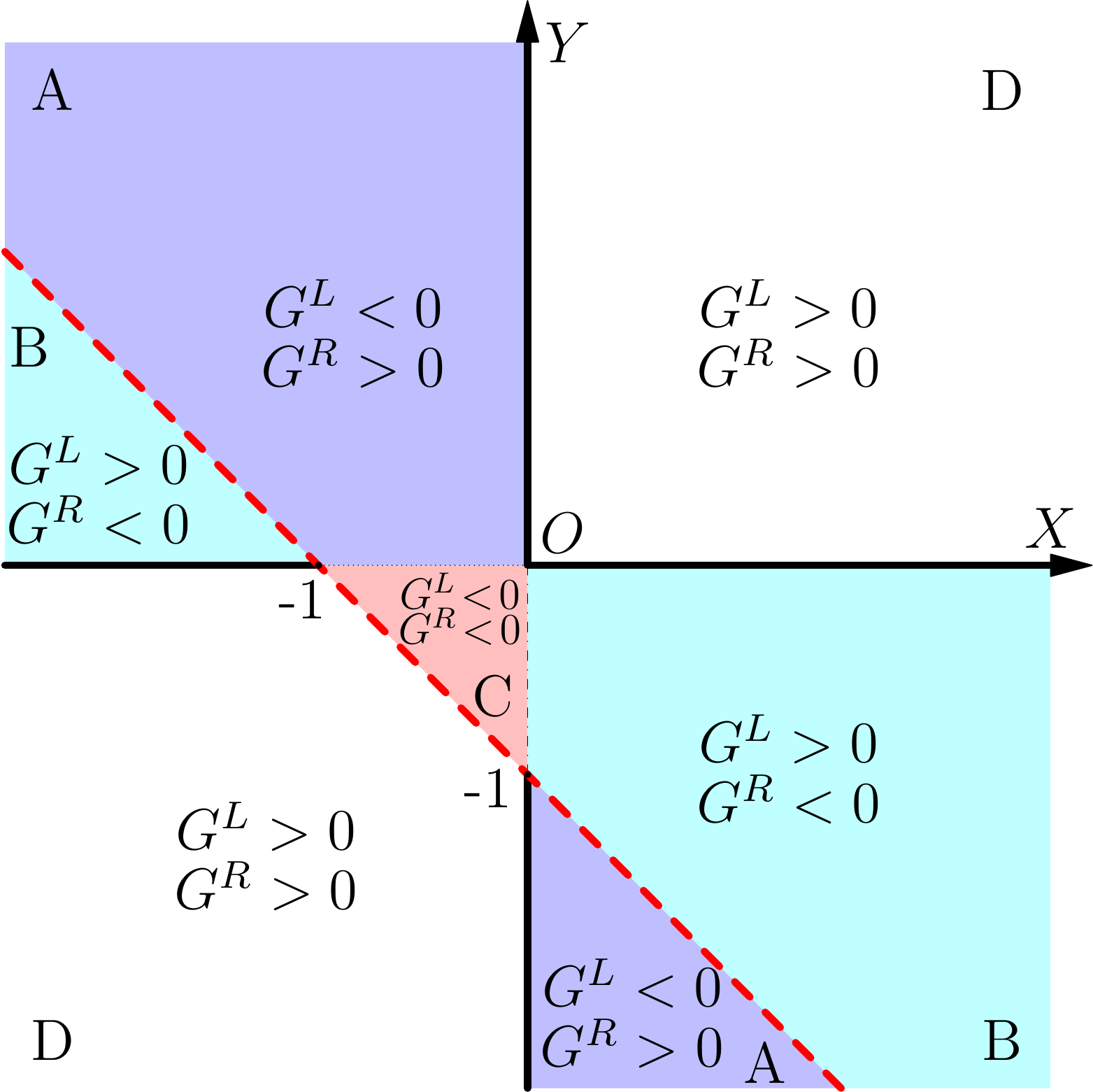}\\
  \caption{The phase diagram on the $X-Y$ plane that present NDTC.}
  \label{fig:xy}
\end{figure}

The blue and cyan shaded areas A and B in Fig.~\ref{fig:xy} are not bounded on the $X$-$Y$ plane. They correspond to one of $G^L$ and $G^R$ is negative and the other is positive, i.e., $G^LG^R<0$ which is equivalent to
\begin{equation}
XY<0.
\label{NDTC condition 0}
\end{equation}
Eq.~(\ref{NDTC condition 0}) includes the situation that $G^{L(R)}$ is infinite (on the thick solid red lines inside the second and the fourth quarters of the $X$-$Y$ plane). We can write Eq.~(\ref{NDTC condition 0}) in a more transparent way by rewriting the temperature dependence of the TCR $R^J(T_1^J,T_2^J)=R^J(\hat T^J,\bar T^J)$ where $\hat T^J\equiv T_1^J-T_2^J$ and $\bar{T}^J\equiv T_1^J+T_2^J$:
\begin{equation}
\left\vert q\frac{\partial R^J}{\partial\hat T^J}-1\right\vert<\left\vert q\frac{\partial R^J}{\partial\bar T^J}\right\vert.
\label{NDTC condition 0.5}
\end{equation}
Eqs.~(\ref{NDTC condition 0.5}) implies that $\left\vert q\frac{\partial R^J}{\partial\bar T^J}\right\vert>0$, i.e., the TCR must be dependent on $\bar T^J$. This is physically significant, because in thermal transport we have a natural temperature ground of absolute zero temperature. This demonstrates the drastic difference between the thermal and electrical transport, where in the latter case the junction behaviour only depends on the voltage difference (not the average voltage) across the junction.

The red shaded area C in Fig.~\ref{fig:xy} is bounded and it corresponds to the case that both $1-q\partial R^J/\partial T^J_1$ and $1+q\partial R^J/\partial T^J_2$ are negative and $R^D$ is positive, equivalent to
\begin{equation}
q\frac{\partial R^J}{\partial\hat T^J}-1>\left\vert q\frac{\partial R^J}{\partial\bar T^J}\right\vert,
\vert q\vert<\frac{R^J+J_1+J_2}{\left\vert\frac{\partial R^J}{\partial T_1^J}\right\vert J_1+\left\vert\frac{\partial R^J}{\partial T_2^J}\right\vert J_2}.
\label{NDTC condition 1}
\end{equation}

Eq.~(\ref{NDTC condition 0.5}) and Eq.~(\ref{NDTC condition 1}) imply that 
\begin{equation}
\vert q\vert>\left(\text{max}\left\{\left\vert\frac{\partial R^J}{\partial T_1^J}\right\vert,\left\vert\frac{\partial R^J}{\partial T_2^J}\right\vert\right\}\right)^{-1}
\end{equation}
\label{q min bound 1}
and
\begin{equation}
\vert q\vert>\left(\text{min}\left\{\left\vert\frac{\partial R^J}{\partial T_1^J}\right\vert,\left\vert\frac{\partial R^J}{\partial T_2^J}\right\vert\right\}\right)^{-1}
\label{q min bound 2}
\end{equation}
respectively, suggesting a positive minimum on the heat current for the existence of NDTC. This again indicates that the existence of NDTC is in the \emph{nonlinear} regime, beyond low heat current.

At the onset of NDTC, we have either $G^L=0$ or $G^R=0$ and the magnitude of heat current (vs. temperature bias) reaches its local maximum. Taking the case of $G^L=0$ as an example, we have the following variation rates at the vicinity of $G^L=0$ (Appendix B):
\begin{equation}
\begin{aligned}
&\frac{\delta(T_1^J-T_2^J)}{\delta T^L}\!=\!F_1(x^L)\!-\!G^L(J_1\!+\!J_2)\!\approx\! F_1(x^L)>0,\\
&\frac{\delta(T_2^J-T^R)}{\delta T^L}=G^LJ_2.
\end{aligned}
\label{var rates temp}
\end{equation}
When the system enters the NDTC regime of $G^L\lesssim 0$, the temperature increase $\delta T^L$ of $T^L$ is distributed over $[x^L,x^R]$ in such a way that the temperature drop over $[x^J,x^R]$ is decreasing with increasing $T^L$ (which is the manifestation of NDTC) while the temperature drop over the junction is increasing with $T^L$, as shown in Fig.~\ref{fig:Tprof}.
\vspace{0.5cm}

\noindent\textit{Systems with multiple abrupt junctions} can exhibit both NDTC and infinite DTCs, but the detailed conditions for their occurrence are more complicated. It can be proved that NDTC still requires that at least one of the junctions possess intrinsic junction NDTC (Appendix C). Nevertheless, these junctions can be grouped into a single effective junction with its properties determined by the way the junctions are organized (e.g., the order and the connection materials) and by the properties of those individual junctions, as shown in Fig.~\ref{fig:scheme}(c). After identifying the effective TCR $R^J_{\text{eff}}$, we can treat the system with multiple junctions as one with a single junction. The discussions in the previous section can be readily applied by simply replacing $R^J$ with $R^J_{\text{eff}}$.

This procedure also provides us a routine to engineer the TCR. For example, we can construct a system composed of three segments in $[x^L,x_1^J]$, $[x_1^J,x_2^J]$ and $[x_2^J,x^R]$. Suppose that $[x^L,x_1^J]$ and $[x_2^J,x^R]$ contain the same kind of uniform material and the material in $[x_1^J,x_2^J]$ is also uniform but different. We can have a single effective junction with its TCR $R^J_{\text{eff}}=q/(T_1^J-T_2^J)$ where $T_1^J$ is the temperature at $x_1^J$ at the side of $[x^L,x_1^J]$, $T_2^J$ is the temperature at $x_2^J$ at the side of $[x_2^J,x^R]$ and $q$ is the heat current flowing across the effective junction. In this way, we have a symmetrical effective junction, i.e., $R^J_{\text{eff}}(T_1^J,T_2^J)=R^J_{\text{eff}}(T_2^J,T_1^J)$.


\vspace{0.5cm}

In conclusion, we have studied the steady state 1D thermal transport in the diffusive regime without heat sources or sinks. The Fourier's law is applied to calculate the differential thermal conductance. We find that NDTC (in the case that the temperature at one end is fixed) cannot exist in systems without any abrupt thermal junctions. However, we could have NDTC if and only if a junction with intrinsic junction NDTC is introduced. Our predictions provide a theoretical foundation to experimentally realize NDTC through careful thermal contact engineering, though it remains an open question to realize a junction with intrinsic junction NDTC.

\vspace{0.5cm}

This work is partially supported by the Semiconductor Research Corporation (SRC) - Nanoelectronics Research Initiative (NRI) via Midwest Institute for Nanoelectronics Discovery (MIND) and the Cooling Technologies Research Center (CTRC) at Purdue University. JH thanks Prof. Xiulin Ruan (Purdue University) and Dr. Xingpeng Yan for useful discussions.

\section*{Appendix}

\subsection{Systems without abrupt junctions}
To calculate the differential thermal conductance (DTC), we start from the variation of $q=-\kappa(x,T(x))\frac{\mathrm{d}T(x)}{\mathrm{d}x}$:
\begin{equation}
\delta q=-\kappa(x,T)\frac{d}{dx}\delta T-\frac{\partial\kappa}{\partial T}\frac{\dif T}{\dif x}\delta T.
\label{app:heat eq var}
\end{equation}
We define
\begin{equation}
U^{L(R)}(x)\equiv\frac{\delta T(x)}{\delta T^{L(R)}}
\label{app:temp var rate}
\end{equation}
which according to Eq.~(\ref{app:heat eq var}) satisfies
\begin{equation}
\frac{\dif }{\dif x}U^{L(R)}+\frac{1}{\kappa(x,T)}\frac{\partial\kappa}{\partial T}\frac{\dif T}{\dif x}U^{L(R)}+\frac{\eta^{L(R)}G^{L(R)}}{\kappa(x,T)}=0.
\label{app:U eq}
\end{equation}
where $\eta^L=1$ and $\eta^R=-1$. At the boundaries we have
\begin{equation}
\delta T(x^{L(R)})=\delta T^{L(R)}\;\;\text{while}\;\;\delta T(x^{R(L)})=0,
\label{app:heat boundary var}
\end{equation} 
since $T(x^{L(R)})=T^{L(R)}$. Thus from Eq.~(\ref{app:temp var rate}), the boundary conditions for Eq.~(\ref{app:U eq}) are
\begin{equation}
U^{L}(x^L)=1,U^{L}(x^R)=U^{R}(x^L)=0,U^{R}(x^R)=1.
\label{app:U boundary}
\end{equation}
Eq.~(\ref{app:U eq}) is an inhomogeneous linear ordinary differential equation, and the coefficients $\frac{1}{\kappa(x,T)}\frac{\partial\kappa}{\partial T}\frac{\dif T}{\dif x}$ and $\frac{\eta^{L(R)}G^{L(R)}}{\kappa(x,T)}$ are functions of $x$ only, since $T(x)$ is already formally solved from Eq.~(\ref{heat eq}) and boundary conditions Eq.~(\ref{heat boundary}). The solution 
to Eq.~(\ref{app:U eq}) is
\begin{equation}
{
U^{L\!(R)}(x)\!\!=\!\!\frac{U^{L\!(R)}(x^L)}{F(x)}\!\!-\!\!\frac{\eta^{L\!(R)}G^{L\!(R)}}{F(x)}\!\!\int_{x^L}^x\!\frac{F(x^\prime)}{\kappa(x^\prime,T(x^\prime))}\dif x^\prime,
}
\label{app:U general sol}
\end{equation}
where 
\begin{equation}
F(x)\equiv\exp\left(\int_{x^L}^x\frac{1}{\kappa(x^\prime,T(x^\prime))}\frac{\partial\kappa}{\partial T}\frac{\dif T}{\dif x^\prime}\dif x^\prime\right).
\label{app:F}
\end{equation}
By evaluating Eq.~(\ref{app:U general sol}) at $x=x^R$, we obtain
\begin{equation}
G^{L(R)}=J^{-1}\eta^{L(R)}\left[U^{L(R)}(x^L)-F(x^R)U^{L(R)}(x^R)\right],
\label{app:GLR}
\end{equation}
where
\begin{equation}
J\equiv\int_{x^L}^{x^R}\frac{F(x^\prime)}{\kappa(x^\prime,T(x^\prime))}\dif x^\prime.
\label{app:J}
\end{equation}
From Eq.~(\ref{app:U boundary}) and Eq.~(\ref{app:GLR}), we have
\begin{equation}
G^L=J^{-1},\;\;G^R=F(x^R)J^{-1}.
\label{app:GLR2}
\end{equation}

\subsection{Systems with a single junction}
For a system composed of two segments lying in $[x^L,x^J]$ and $[x^J,x^R]$, we denote the temperature profiles $T_1(x)$ and $T_2(x)$ and the thermal conductivity $\kappa_1(x,T_1(x))$ and $\kappa_2(x,T_1(x))$, receptively. If a heat current is flowing in the system, the temperature profiles satisfy
\begin{equation}
\begin{aligned}
&q=-\kappa_1(x,T_1(x))\frac{\dif T_1}{\dif x},\;\;\;\;x^L<x<x^J,\\
&q=-\kappa_2(x,T_2(x))\frac{\dif T_2}{\dif x},\;\;\;\;x^J<x<x^R
\end{aligned}
\label{app:heat current var coupled}
\end{equation}
with the boundary conditions
\begin{equation}
T_1^J-T_2^J=qR^J(T_1^J,T_2^J),T_1(x^L)=T^L,T_2(x^R)=T^R.
\label{app:boundary condition coupled}
\end{equation}
Applying the variation $\delta T^{L(R)}$ of the boundary temperature at one end, the resulting temperature profiles $T_1$ and $T_2$ are varied to $T_1(x)+\delta T_1(x)$ and $T_2(x)+\delta T_2(x)$ respectively. Of course we have
\begin{equation}
\delta T_1(x^L)=\delta T^L,\;\;\delta T_2(x^R)=\delta T^R.
\label{app:boundary temperature variation coupled}
\end{equation}
The heat current $q$ is varied to $q+\delta q$. The junction temperatures $T_1^J$ and $T_2^J$ are varied to $T_1^J+\delta T_1^J$ and $T_2^J+\delta T_2^J$ respectively. We then define the following functions
\begin{equation}
U_1^{L(R)}\equiv\frac{\delta T_1}{\delta T^{L(R)}}\;\;\text{and}\;\;U_2^{L(R)}\equiv\frac{\delta T_2}{\delta T^{L(R)}}
\label{app:U12LR coupled}
\end{equation}
on $[x^L,x^J]$ and $[x^J,x^R]$, respectively. They satisfy the following equations
\begin{equation}
\eta^{\!L\!(R)}\!G^{L\!(R)}\!=\!-\kappa_i\!\frac{\dif U_i^{L\!(R)}}{\dif x}\!-\!\frac{\partial\kappa_i}{\partial T_i}\!\frac{\dif T_i}{\dif x}U_i^{L\!(R)},\;i=1,2
\label{app:U eq coupled}
\end{equation}
and boundary conditions
\begin{equation}
U_1^{L}(x^L)=1,U_1^{R}(x^L)=U_2^{L}(x^R)=0,U_2^{R}(x^R)=1.
\end{equation}
Their solutions are
\begin{equation}
U_i^{L\!(R)}\!(x)\!\!=\!\!\frac{U_i^{L\!(R)}\!(x^J)}{F_i(x)}\!-\!\frac{\eta^{L\!(R)}G^{L\!(R)}}{F_i(x)}\!\!\int_{x^J}^x\frac{F_i(x^\prime)}{\kappa_i(x^\prime,T_i(x^\prime))}\dif x^\prime
\label{app:U general sol coupled 1}
\end{equation}
where 
\begin{equation}
F_i(x)\!\equiv\!\exp\!\left(\int_{x^J}^x\frac{1}{\kappa_i(x^\prime,T_i(x^\prime))}\frac{\partial\kappa_i}{\partial T_i}\frac{\dif T_i}{\dif x^\prime}\dif x^\prime\!\right),\;i=1,2.
\label{app:F coupled 2}
\end{equation}
By evaluating Eq.~(\ref{app:U general sol coupled 1}) at $x^L$ for $i=1$ and at $x^R$ for $i=2$, we have
\begin{equation}
\begin{aligned}
U_1^{L(R)}(x^J)=&F_1(x^L)U_1^{L(R)}(x^L)\\
&+\eta^{L\!(R)}G^{L\!(R)}\int_{x^J}^{x^L}\frac{F_1(x^\prime)}{\kappa_1(x^\prime,T_1(x^\prime))}\dif x^\prime,\\
U_2^{L(R)}(x^J)=&F_2(x^R)U_2^{L(R)}(x^R)\\
&+\eta^{L\!(R)}G^{L\!(R)}\int_{x^J}^{x^R}\frac{F_2(x^\prime)}{\kappa_2(x^\prime,T_2(x^\prime))}\dif x^\prime.
\end{aligned}
\label{app:ULRxJ}
\end{equation}
The variation of the first equation in Eq.~(\ref{app:boundary condition coupled}) gives
\begin{equation}
\begin{aligned}
U_1^{L(R)}&(x^J)-U_2^{L(R)}(x^J)=\eta^{L(R)}G^{L(R)}R^J\\&+q\frac{\partial R^J}{\partial T_1^J}U_1^{L(R)}(x^J)+q\frac{\partial R^J}{\partial T_2^J}U_2^{L(R)}(x^J).
\end{aligned}
\label{app:U junction var coupled}
\end{equation}
By inserting Eq.~(\ref{app:ULRxJ}) into Eq.~(\ref{app:U junction var coupled}), we finally get the DTC
\begin{equation}
\begin{aligned}
G^L=\left(1-q\frac{\partial R^J}{\partial T_1^J}\right)\frac{F_1(x^L)}{R^D},\\
G^R=\left(1+q\frac{\partial R^J}{\partial T_2^J}\right)\frac{F_2(x^R)}{R^D},
\end{aligned}
\label{app:GLR coupled}
\end{equation}
where
\begin{equation}
\begin{aligned}
R^D\equiv R^J+\displaystyle\left(1-q\frac{\partial R^J}{\partial T_1^J}\right)J_1+\left(1+q\frac{\partial R^J}{\partial T_2^J}\right)J_2,
\end{aligned}
\label{app:RD}
\end{equation}
with
\begin{equation}
\begin{aligned}
J_1\!\!\equiv\!\!\!\int_{x^L}^{x^J}\!\!\!\frac{F_1(x^\prime)}{\kappa_1\!(x^\prime\!,T_1\!(x^\prime))}\dif x^\prime,
J_2\!\!\equiv\!\!\!\int_{x^J}^{x^R}\!\!\!\frac{F_2(x^\prime)}{\kappa_2\!(x^\prime\!,T_2\!(x^\prime))}\dif x^\prime.
\end{aligned}
\end{equation}

From Eq.~(\ref{app:ULRxJ}), we can also readily write down
\begin{equation}
\begin{aligned}
\frac{\delta(T_1^J-T_2^J)}{\delta T^L}&=U_1^L(x^J)-U_2^L(x^J)\\
&=F_1(x^L)-G^L(J_1+J_2),\\
\frac{\delta(T_2^J-T^R)}{\delta T^L}&=U_2^L(x^J)=G^LJ_2.
\end{aligned}
\end{equation}

\subsection{Existence of NDTC for systems with multiple junctions}

To prove that the existence of NDTC for systems with multiple junctions requires that at least one of the junctions has intrinsic junction NDTC, we prove the converse by induction, i.e., there is no NDTC if none of the junctions has intrinsic junction NDTC.

Assume that the system lying in $[x^J,x^R]$ contains an arbitrary number of junctions and the DTCs for this system are non-negative. We now add a new segment (with no junctions within the segment) lying in $[x^L,x^J]$ to the existing system. We denote $T^{L(R)}$ as the temperature at $x^{L(R)}$. We assume that the new junction at $x^J$ with TCR $R^J(T_1^J,T_2^J)$ has no intrinsic junction NDTC. Here $T_1^J$ and $T_2^J$ are the temperatures at $x^J$ at the side of $[x^L,x^J]$ and $[x^J,x^R]$ respectively. Suppose we raise the temperature $T^L$ infinitesimally to $T^L+\delta T^L$ ($\delta T^L>0$) while keeping $T^R$ fixed. The temperature $T_{1(2)}^J$ is then varied to $T_{1(2)}^J+\delta T_{1(2)}^J$ and the heat current is changed from $q$ to $q+\delta q$. At the junction, we have $\delta q=G_1^J\delta T_1^J-G_2^J\delta T_2^J$ where $G_{1(2)}^J\ge 0$ are the junction DTCs. On the other hand, the system in $[x^J,x^R]$ has non-negative DTCs, i.e., $\delta q=G^B\delta T_2^J$ where $G^B\ge 0$. We thus have $G_1^J\delta T_1^J=(G_2^J+G^B)\delta T_2^J$. Of course $\delta T_1^J$ cannot be negative. 
If either $G_2^J$ or $G^B$ is positive, we will have $\delta T_2^J=G_1^J\delta T_1^J/(G_2^J+G^B)\ge 0$ and thus $\delta q=G^B\delta T_2^J\ge 0$ and there is no NDTC. If both $G_2^J$ and $G^B$ are zero, we will have $\delta q=G_1^J\delta T_1^J\ge 0$ and there is no NDTC.



\begin{thebibliography}{32}%
\makeatletter
\providecommand \@ifxundefined [1]{%
 \@ifx{#1\undefined}
}%
\providecommand \@ifnum [1]{%
 \ifnum #1\expandafter \@firstoftwo
 \else \expandafter \@secondoftwo
 \fi
}%
\providecommand \@ifx [1]{%
 \ifx #1\expandafter \@firstoftwo
 \else \expandafter \@secondoftwo
 \fi
}%
\providecommand \natexlab [1]{#1}%
\providecommand \enquote  [1]{``#1''}%
\providecommand \bibnamefont  [1]{#1}%
\providecommand \bibfnamefont [1]{#1}%
\providecommand \citenamefont [1]{#1}%
\providecommand \href@noop [0]{\@secondoftwo}%
\providecommand \href [0]{\begingroup \@sanitize@url \@href}%
\providecommand \@href[1]{\@@startlink{#1}\@@href}%
\providecommand \@@href[1]{\endgroup#1\@@endlink}%
\providecommand \@sanitize@url [0]{\catcode `\\12\catcode `\$12\catcode
  `\&12\catcode `\#12\catcode `\^12\catcode `\_12\catcode `\%12\relax}%
\providecommand \@@startlink[1]{}%
\providecommand \@@endlink[0]{}%
\providecommand \url  [0]{\begingroup\@sanitize@url \@url }%
\providecommand \@url [1]{\endgroup\@href {#1}{\urlprefix }}%
\providecommand \urlprefix  [0]{URL }%
\providecommand \Eprint [0]{\href }%
\providecommand \doibase [0]{http://dx.doi.org/}%
\providecommand \selectlanguage [0]{\@gobble}%
\providecommand \bibinfo  [0]{\@secondoftwo}%
\providecommand \bibfield  [0]{\@secondoftwo}%
\providecommand \translation [1]{[#1]}%
\providecommand \BibitemOpen [0]{}%
\providecommand \bibitemStop [0]{}%
\providecommand \bibitemNoStop [0]{.\EOS\space}%
\providecommand \EOS [0]{\spacefactor3000\relax}%
\providecommand \BibitemShut  [1]{\csname bibitem#1\endcsname}%
\let\auto@bib@innerbib\@empty
\bibitem [{\citenamefont {Lepri}\ \emph {et~al.}(2003)\citenamefont {Lepri},
  \citenamefont {Livi},\ and\ \citenamefont {Politi}}]{Lepri03}%
  \BibitemOpen
  \bibfield  {author} {\bibinfo {author} {\bibfnamefont {S.}~\bibnamefont
  {Lepri}}, \bibinfo {author} {\bibfnamefont {R.}~\bibnamefont {Livi}}, \ and\
  \bibinfo {author} {\bibfnamefont {A.}~\bibnamefont {Politi}},\ }\href@noop {}
  {\bibfield  {journal} {\bibinfo  {journal} {Phys. Rep.}\ }\textbf {\bibinfo
  {volume} {377}},\ \bibinfo {pages} {1} (\bibinfo {year} {2003})}\BibitemShut
  {NoStop}%
\bibitem [{\citenamefont {Li}\ \emph {et~al.}(2012)\citenamefont {Li},
  \citenamefont {Ren}, \citenamefont {Wang}, \citenamefont {Zhang},
  \citenamefont {H\"anggi},\ and\ \citenamefont {Li}}]{RevModPhys.84.1045}%
  \BibitemOpen
  \bibfield  {author} {\bibinfo {author} {\bibfnamefont {N.}~\bibnamefont
  {Li}}, \bibinfo {author} {\bibfnamefont {J.}~\bibnamefont {Ren}}, \bibinfo
  {author} {\bibfnamefont {L.}~\bibnamefont {Wang}}, \bibinfo {author}
  {\bibfnamefont {G.}~\bibnamefont {Zhang}}, \bibinfo {author} {\bibfnamefont
  {P.}~\bibnamefont {H\"anggi}}, \ and\ \bibinfo {author} {\bibfnamefont
  {B.}~\bibnamefont {Li}},\ }\href {\doibase 10.1103/RevModPhys.84.1045}
  {\bibfield  {journal} {\bibinfo  {journal} {Rev. Mod. Phys.}\ }\textbf
  {\bibinfo {volume} {84}},\ \bibinfo {pages} {1045} (\bibinfo {year}
  {2012})}\BibitemShut {NoStop}%
\bibitem [{\citenamefont {Terraneo}\ \emph {et~al.}(2002)\citenamefont
  {Terraneo}, \citenamefont {Peyrard},\ and\ \citenamefont
  {Casati}}]{PhysRevLett.88.094302}%
  \BibitemOpen
  \bibfield  {author} {\bibinfo {author} {\bibfnamefont {M.}~\bibnamefont
  {Terraneo}}, \bibinfo {author} {\bibfnamefont {M.}~\bibnamefont {Peyrard}}, \
  and\ \bibinfo {author} {\bibfnamefont {G.}~\bibnamefont {Casati}},\
  }\href@noop {} {\bibfield  {journal} {\bibinfo  {journal} {Phys. Rev. Lett.}\
  }\textbf {\bibinfo {volume} {88}},\ \bibinfo {pages} {094302} (\bibinfo
  {year} {2002})}\BibitemShut {NoStop}%
\bibitem [{\citenamefont {Eckmann}\ and\ \citenamefont
  {Mej\'{i}a-Monasterio}(2006)}]{PhysRevLett.97.094301}%
  \BibitemOpen
  \bibfield  {author} {\bibinfo {author} {\bibfnamefont {J.-P.}\ \bibnamefont
  {Eckmann}}\ and\ \bibinfo {author} {\bibfnamefont {C.}~\bibnamefont
  {Mej\'{i}a-Monasterio}},\ }\href@noop {} {\bibfield  {journal} {\bibinfo
  {journal} {Phys. Rev. Lett.}\ }\textbf {\bibinfo {volume} {97}},\ \bibinfo
  {pages} {094301} (\bibinfo {year} {2006})}\BibitemShut {NoStop}%
\bibitem [{\citenamefont {Li}\ \emph {et~al.}(2004)\citenamefont {Li},
  \citenamefont {Wang},\ and\ \citenamefont {Casati}}]{Li04}%
  \BibitemOpen
  \bibfield  {author} {\bibinfo {author} {\bibfnamefont {B.}~\bibnamefont
  {Li}}, \bibinfo {author} {\bibfnamefont {L.}~\bibnamefont {Wang}}, \ and\
  \bibinfo {author} {\bibfnamefont {G.}~\bibnamefont {Casati}},\ }\href@noop {}
  {\bibfield  {journal} {\bibinfo  {journal} {Phys. Rev. Lett.}\ }\textbf
  {\bibinfo {volume} {93}},\ \bibinfo {pages} {184301} (\bibinfo {year}
  {2004})}\BibitemShut {NoStop}%
\bibitem [{\citenamefont {Casati}\ \emph {et~al.}(2007)\citenamefont {Casati},
  \citenamefont {Mej\'{i}a-Monasterio},\ and\ \citenamefont
  {Prosen}}]{PhysRevLett.98.104302}%
  \BibitemOpen
  \bibfield  {author} {\bibinfo {author} {\bibfnamefont {G.}~\bibnamefont
  {Casati}}, \bibinfo {author} {\bibfnamefont {C.}~\bibnamefont
  {Mej\'{i}a-Monasterio}}, \ and\ \bibinfo {author} {\bibfnamefont
  {T.}~\bibnamefont {Prosen}},\ }\href@noop {} {\bibfield  {journal} {\bibinfo
  {journal} {Phys. Rev. Lett.}\ }\textbf {\bibinfo {volume} {98}},\ \bibinfo
  {pages} {104302} (\bibinfo {year} {2007})}\BibitemShut {NoStop}%
\bibitem [{\citenamefont {Peyrard}(2006)}]{Peyrard06}%
  \BibitemOpen
  \bibfield  {author} {\bibinfo {author} {\bibfnamefont {M.}~\bibnamefont
  {Peyrard}},\ }\href@noop {} {\bibfield  {journal} {\bibinfo  {journal}
  {Europhysics Lett.}\ }\textbf {\bibinfo {volume} {76}},\ \bibinfo {pages}
  {49} (\bibinfo {year} {2006})}\BibitemShut {NoStop}%
\bibitem [{\citenamefont {Chang}\ \emph {et~al.}(2006)\citenamefont {Chang},
  \citenamefont {Okawa}, \citenamefont {Majumdar},\ and\ \citenamefont
  {Zettl}}]{Chang06}%
  \BibitemOpen
  \bibfield  {author} {\bibinfo {author} {\bibfnamefont {C.~W.}\ \bibnamefont
  {Chang}}, \bibinfo {author} {\bibfnamefont {D.}~\bibnamefont {Okawa}},
  \bibinfo {author} {\bibfnamefont {A.}~\bibnamefont {Majumdar}}, \ and\
  \bibinfo {author} {\bibfnamefont {A.}~\bibnamefont {Zettl}},\ }\href@noop {}
  {\bibfield  {journal} {\bibinfo  {journal} {Science}\ }\textbf {\bibinfo
  {volume} {314}},\ \bibinfo {pages} {1121} (\bibinfo {year}
  {2006})}\BibitemShut {NoStop}%
\bibitem [{\citenamefont {Yang}\ \emph {et~al.}(2007)\citenamefont {Yang},
  \citenamefont {Li}, \citenamefont {Wang},\ and\ \citenamefont {Li}}]{Yang07}%
  \BibitemOpen
  \bibfield  {author} {\bibinfo {author} {\bibfnamefont {N.}~\bibnamefont
  {Yang}}, \bibinfo {author} {\bibfnamefont {N.}~\bibnamefont {Li}}, \bibinfo
  {author} {\bibfnamefont {L.}~\bibnamefont {Wang}}, \ and\ \bibinfo {author}
  {\bibfnamefont {B.}~\bibnamefont {Li}},\ }\href@noop {} {\bibfield  {journal}
  {\bibinfo  {journal} {Phys. Rev. B}\ }\textbf {\bibinfo {volume} {76}},\
  \bibinfo {pages} {020301} (\bibinfo {year} {2007})}\BibitemShut {NoStop}%
\bibitem [{\citenamefont {Hu}\ \emph {et~al.}(2009)\citenamefont {Hu},
  \citenamefont {Ruan},\ and\ \citenamefont {Chen}}]{Hu09}%
  \BibitemOpen
  \bibfield  {author} {\bibinfo {author} {\bibfnamefont {J.}~\bibnamefont
  {Hu}}, \bibinfo {author} {\bibfnamefont {X.}~\bibnamefont {Ruan}}, \ and\
  \bibinfo {author} {\bibfnamefont {Y.~P.}\ \bibnamefont {Chen}},\ }\href@noop
  {} {\bibfield  {journal} {\bibinfo  {journal} {Nano Lett.}\ }\textbf
  {\bibinfo {volume} {9}},\ \bibinfo {pages} {2730} (\bibinfo {year}
  {2009})}\BibitemShut {NoStop}%
\bibitem [{\citenamefont {Pereira}(2010)}]{Pereira10}%
  \BibitemOpen
  \bibfield  {author} {\bibinfo {author} {\bibfnamefont {E.}~\bibnamefont
  {Pereira}},\ }\href@noop {} {\bibfield  {journal} {\bibinfo  {journal} {Phys.
  Rev. E}\ }\textbf {\bibinfo {volume} {82}},\ \bibinfo {pages} {040101}
  (\bibinfo {year} {2010})}\BibitemShut {NoStop}%
\bibitem [{\citenamefont {Sawaki}\ \emph {et~al.}(2011)\citenamefont {Sawaki},
  \citenamefont {Kobayashi}, \citenamefont {Moritomo},\ and\ \citenamefont
  {Terasaki}}]{Sawaki11}%
  \BibitemOpen
  \bibfield  {author} {\bibinfo {author} {\bibfnamefont {D.}~\bibnamefont
  {Sawaki}}, \bibinfo {author} {\bibfnamefont {W.}~\bibnamefont {Kobayashi}},
  \bibinfo {author} {\bibfnamefont {Y.}~\bibnamefont {Moritomo}}, \ and\
  \bibinfo {author} {\bibfnamefont {I.}~\bibnamefont {Terasaki}},\ }\href@noop
  {} {\bibfield  {journal} {\bibinfo  {journal} {Appl. Phys. Lett.}\ }\textbf
  {\bibinfo {volume} {98}},\ \bibinfo {pages} {081915} (\bibinfo {year}
  {2011})}\BibitemShut {NoStop}%
\bibitem [{\citenamefont {Go}\ and\ \citenamefont {Sen}(2010)}]{Go10}%
  \BibitemOpen
  \bibfield  {author} {\bibinfo {author} {\bibfnamefont {D.~B.}\ \bibnamefont
  {Go}}\ and\ \bibinfo {author} {\bibfnamefont {M.}~\bibnamefont {Sen}},\
  }\href@noop {} {\bibfield  {journal} {\bibinfo  {journal} {J. Heat Transfer}\
  }\textbf {\bibinfo {volume} {132}},\ \bibinfo {pages} {124502} (\bibinfo
  {year} {2010})}\BibitemShut {NoStop}%
\bibitem [{\citenamefont {Dames}(2009)}]{Dames09}%
  \BibitemOpen
  \bibfield  {author} {\bibinfo {author} {\bibfnamefont {C.}~\bibnamefont
  {Dames}},\ }\href@noop {} {\bibfield  {journal} {\bibinfo  {journal} {J. Heat
  Transfer}\ }\textbf {\bibinfo {volume} {131}},\ \bibinfo {eid} {061301}
  (\bibinfo {year} {2009})}\BibitemShut {NoStop}%
\bibitem [{\citenamefont {Li}\ \emph {et~al.}(2006)\citenamefont {Li},
  \citenamefont {Wang},\ and\ \citenamefont {Casati}}]{Li06}%
  \BibitemOpen
  \bibfield  {author} {\bibinfo {author} {\bibfnamefont {B.}~\bibnamefont
  {Li}}, \bibinfo {author} {\bibfnamefont {L.}~\bibnamefont {Wang}}, \ and\
  \bibinfo {author} {\bibfnamefont {G.}~\bibnamefont {Casati}},\ }\href
  {\doibase 10.1063/1.2191730} {\bibfield  {journal} {\bibinfo  {journal}
  {Appl. Phys. Lett.}\ }\textbf {\bibinfo {volume} {88}},\ \bibinfo {eid}
  {143501} (\bibinfo {year} {2006})}\BibitemShut {NoStop}%
\bibitem [{\citenamefont {Wang}\ and\ \citenamefont {Li}(2007)}]{Wang07}%
  \BibitemOpen
  \bibfield  {author} {\bibinfo {author} {\bibfnamefont {L.}~\bibnamefont
  {Wang}}\ and\ \bibinfo {author} {\bibfnamefont {B.}~\bibnamefont {Li}},\
  }\href@noop {} {\bibfield  {journal} {\bibinfo  {journal} {Phys. Rev. Lett.}\
  }\textbf {\bibinfo {volume} {99}},\ \bibinfo {pages} {177208} (\bibinfo
  {year} {2007})}\BibitemShut {NoStop}%
\bibitem [{\citenamefont {He}\ \emph {et~al.}(2009)\citenamefont {He},
  \citenamefont {Buyukdagli},\ and\ \citenamefont {Hu}}]{He09}%
  \BibitemOpen
  \bibfield  {author} {\bibinfo {author} {\bibfnamefont {D.}~\bibnamefont
  {He}}, \bibinfo {author} {\bibfnamefont {S.}~\bibnamefont {Buyukdagli}}, \
  and\ \bibinfo {author} {\bibfnamefont {B.}~\bibnamefont {Hu}},\ }\href@noop
  {} {\bibfield  {journal} {\bibinfo  {journal} {Phys. Rev. B}\ }\textbf
  {\bibinfo {volume} {80}},\ \bibinfo {pages} {104302} (\bibinfo {year}
  {2009})}\BibitemShut {NoStop}%
\bibitem [{\citenamefont {Zhong}\ \emph {et~al.}(2009)\citenamefont {Zhong},
  \citenamefont {Yang}, \citenamefont {Ai}, \citenamefont {Shao},\ and\
  \citenamefont {Hu}}]{Zhong09}%
  \BibitemOpen
  \bibfield  {author} {\bibinfo {author} {\bibfnamefont {W.-R.}\ \bibnamefont
  {Zhong}}, \bibinfo {author} {\bibfnamefont {P.}~\bibnamefont {Yang}},
  \bibinfo {author} {\bibfnamefont {B.-Q.}\ \bibnamefont {Ai}}, \bibinfo
  {author} {\bibfnamefont {Z.-G.}\ \bibnamefont {Shao}}, \ and\ \bibinfo
  {author} {\bibfnamefont {B.}~\bibnamefont {Hu}},\ }\href@noop {} {\bibfield
  {journal} {\bibinfo  {journal} {Phys. Rev. E}\ }\textbf {\bibinfo {volume}
  {79}},\ \bibinfo {pages} {050103} (\bibinfo {year} {2009})}\BibitemShut
  {NoStop}%
\bibitem [{\citenamefont {Shao}\ \emph {et~al.}(2009)\citenamefont {Shao},
  \citenamefont {Yang}, \citenamefont {Chan},\ and\ \citenamefont
  {Hu}}]{Shao09}%
  \BibitemOpen
  \bibfield  {author} {\bibinfo {author} {\bibfnamefont {Z.-G.}\ \bibnamefont
  {Shao}}, \bibinfo {author} {\bibfnamefont {L.}~\bibnamefont {Yang}}, \bibinfo
  {author} {\bibfnamefont {H.-K.}\ \bibnamefont {Chan}}, \ and\ \bibinfo
  {author} {\bibfnamefont {B.}~\bibnamefont {Hu}},\ }\href@noop {} {\bibfield
  {journal} {\bibinfo  {journal} {Phys. Rev. E}\ }\textbf {\bibinfo {volume}
  {79}},\ \bibinfo {pages} {061119} (\bibinfo {year} {2009})}\BibitemShut
  {NoStop}%
\bibitem [{\citenamefont {He}\ \emph {et~al.}(2010)\citenamefont {He},
  \citenamefont {Ai}, \citenamefont {Chan},\ and\ \citenamefont {Hu}}]{He10}%
  \BibitemOpen
  \bibfield  {author} {\bibinfo {author} {\bibfnamefont {D.}~\bibnamefont
  {He}}, \bibinfo {author} {\bibfnamefont {B.-Q.}\ \bibnamefont {Ai}}, \bibinfo
  {author} {\bibfnamefont {H.-K.}\ \bibnamefont {Chan}}, \ and\ \bibinfo
  {author} {\bibfnamefont {B.}~\bibnamefont {Hu}},\ }\href@noop {} {\bibfield
  {journal} {\bibinfo  {journal} {Phys. Rev. E}\ }\textbf {\bibinfo {volume}
  {81}},\ \bibinfo {pages} {041131} (\bibinfo {year} {2010})}\BibitemShut
  {NoStop}%
\bibitem [{\citenamefont {Ai}\ and\ \citenamefont {Hu}(2011)}]{Ai11}%
  \BibitemOpen
  \bibfield  {author} {\bibinfo {author} {\bibfnamefont {B.-Q.}\ \bibnamefont
  {Ai}}\ and\ \bibinfo {author} {\bibfnamefont {B.}~\bibnamefont {Hu}},\
  }\href@noop {} {\bibfield  {journal} {\bibinfo  {journal} {Phys. Rev. E}\
  }\textbf {\bibinfo {volume} {83}},\ \bibinfo {pages} {011131} (\bibinfo
  {year} {2011})}\BibitemShut {NoStop}%
\bibitem [{\citenamefont {Ai}\ \emph {et~al.}(2011)\citenamefont {Ai},
  \citenamefont {Zhong},\ and\ \citenamefont {Hu}}]{Ai11b}%
  \BibitemOpen
  \bibfield  {author} {\bibinfo {author} {\bibfnamefont {B.-Q.}\ \bibnamefont
  {Ai}}, \bibinfo {author} {\bibfnamefont {W.-R.}\ \bibnamefont {Zhong}}, \
  and\ \bibinfo {author} {\bibfnamefont {B.}~\bibnamefont {Hu}},\ }\href@noop
  {} {\bibfield  {journal} {\bibinfo  {journal} {Phys. Rev. E}\ }\textbf
  {\bibinfo {volume} {83}},\ \bibinfo {pages} {052102} (\bibinfo {year}
  {2011})}\BibitemShut {NoStop}%
\bibitem [{\citenamefont {Hu}\ \emph {et~al.}(2011)\citenamefont {Hu},
  \citenamefont {Wang}, \citenamefont {Vallabhaneni}, \citenamefont {Ruan},\
  and\ \citenamefont {Chen}}]{Hu11}%
  \BibitemOpen
  \bibfield  {author} {\bibinfo {author} {\bibfnamefont {J.}~\bibnamefont
  {Hu}}, \bibinfo {author} {\bibfnamefont {Y.}~\bibnamefont {Wang}}, \bibinfo
  {author} {\bibfnamefont {A.}~\bibnamefont {Vallabhaneni}}, \bibinfo {author}
  {\bibfnamefont {X.}~\bibnamefont {Ruan}}, \ and\ \bibinfo {author}
  {\bibfnamefont {Y.~P.}\ \bibnamefont {Chen}},\ }\href@noop {} {\bibfield
  {journal} {\bibinfo  {journal} {Appl. Phys. Lett.}\ }\textbf {\bibinfo
  {volume} {99}},\ \bibinfo {pages} {113101} (\bibinfo {year}
  {2011})}\BibitemShut {NoStop}%
\bibitem [{\citenamefont {Shao}\ and\ \citenamefont {Yang}(2011)}]{Shao11}%
  \BibitemOpen
  \bibfield  {author} {\bibinfo {author} {\bibfnamefont {Z.-G.}\ \bibnamefont
  {Shao}}\ and\ \bibinfo {author} {\bibfnamefont {L.}~\bibnamefont {Yang}},\
  }\href@noop {} {\bibfield  {journal} {\bibinfo  {journal} {Europhysics
  Lett.}\ }\textbf {\bibinfo {volume} {94}},\ \bibinfo {pages} {34004}
  (\bibinfo {year} {2011})}\BibitemShut {NoStop}%
\bibitem [{\citenamefont {Zhu}\ \emph {et~al.}(2012)\citenamefont {Zhu},
  \citenamefont {Otey},\ and\ \citenamefont {Fan}}]{zhu044104}%
  \BibitemOpen
  \bibfield  {author} {\bibinfo {author} {\bibfnamefont {L.}~\bibnamefont
  {Zhu}}, \bibinfo {author} {\bibfnamefont {C.~R.}\ \bibnamefont {Otey}}, \
  and\ \bibinfo {author} {\bibfnamefont {S.}~\bibnamefont {Fan}},\ }\href@noop
  {} {\bibfield  {journal} {\bibinfo  {journal} {Appl. Phys. Lett.}\ }\textbf
  {\bibinfo {volume} {100}},\ \bibinfo {pages} {044104} (\bibinfo {year}
  {2012})}\BibitemShut {NoStop}%
\bibitem [{\citenamefont {Segal}(2006)}]{Segal06}%
  \BibitemOpen
  \bibfield  {author} {\bibinfo {author} {\bibfnamefont {D.}~\bibnamefont
  {Segal}},\ }\href@noop {} {\bibfield  {journal} {\bibinfo  {journal} {Phys.
  Rev. B}\ }\textbf {\bibinfo {volume} {73}},\ \bibinfo {pages} {205415}
  (\bibinfo {year} {2006})}\BibitemShut {NoStop}%
\bibitem [{\citenamefont {Wu}\ \emph {et~al.}(2009)\citenamefont {Wu},
  \citenamefont {Yu},\ and\ \citenamefont {Segal}}]{PRE.80.041103}%
  \BibitemOpen
  \bibfield  {author} {\bibinfo {author} {\bibfnamefont {L.-A.}\ \bibnamefont
  {Wu}}, \bibinfo {author} {\bibfnamefont {C.~X.}\ \bibnamefont {Yu}}, \ and\
  \bibinfo {author} {\bibfnamefont {D.}~\bibnamefont {Segal}},\ }\href@noop {}
  {\bibfield  {journal} {\bibinfo  {journal} {Phys. Rev. E}\ }\textbf {\bibinfo
  {volume} {80}},\ \bibinfo {pages} {041103} (\bibinfo {year}
  {2009})}\BibitemShut {NoStop}%
\bibitem [{\citenamefont {Walter}\ and\ \citenamefont
  {Thompson}(1998)}]{Walter98}%
  \BibitemOpen
  \bibfield  {author} {\bibinfo {author} {\bibfnamefont {W.}~\bibnamefont
  {Walter}}\ and\ \bibinfo {author} {\bibfnamefont {R.}~\bibnamefont
  {Thompson}},\ }\href@noop {} {\emph {\bibinfo {title} {Ordinary Differential
  Equations}}}\ (\bibinfo  {publisher} {Springer},\ \bibinfo {year}
  {1998})\BibitemShut {NoStop}%
\bibitem [{not({\natexlab{a}})}]{note1}%
  \BibitemOpen
  \href@noop {} {} \bibinfo {note} {If the temperatures at
  both ends vary simultaneously and depend on a parameter $u$ (i.e.,
  $T^{L(R)}(u)$), the DTC is $G=\frac{G^LdT^L/du-G^RdT^R/du}{dT^L/du-dT^R/du}$.
  It is possible to have the numerator of $G$ to be negative while its
  denominator is positive, if the form of $T^{L(R)}(u)$ is designed
  carefully.}\BibitemShut {Stop}%
\bibitem [{not({\natexlab{b}})}]{note2}%
  \BibitemOpen
  \href@noop {} {} \bibinfo {note} {Usually
  $R^J(T_1^J,T_2^J)$ is reduced to $R^J(T^J)$ with $T^J=(T_1^J+T_2^J)/2$ when
  $\vert T_1^J-T_2^J\vert \ll T^J$, as being studied in many experiments (E. T.
  Swartz and R. O. Pohl, Rev. Mod. Phys. 61, 605 (1989)). However, that form of
  $R^J$ is not applicable to the cases involving large currents that we are
  particularly interested in here.}\BibitemShut {Stop}%
\bibitem [{not({\natexlab{c}})}]{note3}%
  \BibitemOpen
  \href@noop {} {} \bibinfo {note} {To have NDTC (negative $G^L$ or $G^R$), it is necessary that at least one of $X$ and $Y$ is negative. Conversely, if at least one of $X$ and $Y$ is negative, since both $X$ and $Y$ (assumed to be continuous with temperature bias) would be positive in the limit of vanishing temperature bias, there exists a critical temperature bias below which both $X$ and $Y$ are positive and slightly above which at least one of $X$ and $Y$ is negative and $\vert X\vert\ll 1$ and $\vert Y\vert\ll 1$, thus one of $G^L$ and $G^R$ must be negative ($1+X+Y$ is positive) and there exists NDTC.}\BibitemShut {Stop}%
\bibitem [{not({\natexlab{d}})}]{note4}%
  \BibitemOpen
  \href@noop {} {} \bibinfo {note} {The zeros of $G^{L}$ and
  $G^R$ exist at $q=[\partial R^J/\partial T_1^J]^{-1}$ and $q=-[\partial
  R^J/\partial T_2^J]^{-1}$ respectively. The zeros may accidentally cancel the
  singularity at $(X,Y)=(0,-1)$ or $(X,Y)=(-1,0)$, but this cancellation rarely
  happen, and can be avoided by slightly perturbing the boundary
  temperatures.}\BibitemShut {Stop}%
\end{thebibliography}

\providecommand{\noopsort}[1]{}\providecommand{\singleletter}[1]{#1}%

\end{document}